\documentstyle[prl,preprint,aps]{revtex} 
\begin{document}
\draft
%

\title {\vbox{\centerline{Cellular automata models of traffic flow} 
\vskip2pt\centerline{along a highway containing a junction}}} %
\author{Simon C. Benjamin, Neil F. Johnson} %
\vspace{.05in}
%
\address {Physics Department, Clarendon Laboratory, 
Oxford University\\
Oxford OX1 3PU, U.K.}
\vspace{.05in}
\author{P.M. Hui}
\vspace{.05in}
%
\address {Physics Department, Chinese University of Hong Kong\\ 
Shatin, New Territories, Hong Kong}
%
\maketitle

\begin{abstract}

We examine various realistic generalizations of the basic cellular
automaton model
describing traffic flow along a highway. In particular, we introduce a {\em
slow-to-start} rule which simulates a possible delay before a car pulls
away from
being stationary. Having discussed the case of a bare highway, we then
consider the
presence of a junction. We study the effects of  acceleration,
disorderness, and
slow-to-start behavior on the queue length at the entrance to the
highway. Interestingly, the junction's efficiency is {\it improved} by
introducing disorderness along the highway, and by imposing a speed
limit.

\end{abstract} \vspace{1.0in}


\narrowtext

Cellular automaton (CA) models have become increasingly
important in the study of traffic flow. 
Although based on a very
simple set of rules, these models are remarkable in their ability to
simulate both gross and subtle features of real traffic. 
In one-dimension (1D), Nagel and Schrekenberg \cite{Nagel} introduced
a stochastic discrete automaton model to study the transition from
laminar traffic flow to start-stop-waves as the car density increases.
Variations on the basic model include introducing separation-dependent
car velocities \cite{Nagatani}, the addition of slower sites and
takeover sites \cite{Chung}, studies of the effect of bottlenecks
\cite{Yukawa} and quenched disorder \cite{Csahok}, etc.. Similar
models have been introduced for traffic flows in more than one lane
\cite{twolane}, in crossings of 1D lanes \cite{Nagatani2} and in
two-dimensions \cite{Biham} with alternate
movements of eastbound and northbound cars simulating the effects
of traffic lights. Effects of inhomogeneities, such as faulty traffic
lights\cite{Chung2} and various mean field theories \cite{Ishibashi},
have also been studied in 2D. 

In this paper we consider traffic flow along a long road, such as a highway,
within the context of a one-dimensional (1D) cellular automaton (CA)
array. We employ the Nagel-Schreckenberg (NS) model[1] and introduce
two modifications: a rule which simulates the disparity between
braking and acceleration, and a set of rules which model a junction
along the highway. 
We begin with a `bare' highway (i.e. no junction) in order to better
understand the model's behavior as the parameters are varied; various
analytic results are given. We then introduce the junction and
study the flow of cars through it as we vary these same parameters.  We
obtain the surprising result that the junction operates more
smoothly when there is a disorderness on the road itself. Junction
performance is also improved by limiting the speed of cars along the
highway.

The basic 1D asymmetric exclusion model is defined on a lattice of
length $N$, with $N$ usually taken to be as large as is computationally
convenient. Each site in the CA lattice has two possible states:
`occupied' by a car and `vacant'. The rule for updating the state at
each site for the most basic model is as follows: all vacant sites
assume the state of sites to their immediate left, and all occupied
sites assume the state of the site to their immediate right. This
implies that cars move to the right if and only if there is a space to
their right. A car may not move into an occupied site even if the
occupying car is moving on in the same step.

The NS model involves two additional rules that produce a
closer simulation of real traffic. (1) Cars may move with a range of
integer speeds, $s=0..s_{max}$. A car with speed $v=s-1$ on the
previous step will move in the next step with a speed given by the
{\em lowest} of the following quantities: (i) $v=s$, (ii) $v=s_{max}$,
or (iii) $v$ equal to the number of vacant sites to its immediate
right. This will be referred to as the {\em acceleration} rule. (2) The
cars are subject to a random disordering effect as follows. For each
car whose scheduled speed for the next update is $v>0$, there is a
probability $P_{fault}$ that it will in fact move with speed $v-1$.
This will be referred to as the {\em disorder} rule. 

The second rule is intended to reflect the flawed behavior of real
(human) drivers. In this spirit, we will now introduce a
further rule, referred to as the {\em slow-to-start} rule. Our rule
will model the small, but finite delay before a car pulls away from
being `static', i.e. when it has reached the head of a queue. This can
arise from a driver's loss of attention as a result of having been
stuck in the queue, or from the slow pick-up of his vehicle's engine.
This rather subtle feature of real traffic is likely to become
important at high car densities, particularly since no such delay is
likely to occur as cars decelerate, i.e. as they brake. The resulting
asymmetry is liable to cause queues to lengthen. We define the
slow-to-start rule as follows: {\em a given static car moves either on
its first opportunity with probability $1-P_{slow}$ or second
opportunity with probability $P_{slow}$}.  We note that the disorder
rule can also cause cars to be slow in moving off from the heads of
queues. However, the disorder rule affects vehicles of all velocities
with equal probability; it introduces a general `noise' into the
system. By contrast, the slow-to-start rule affects only static cars
on the first occasion that they are free to move; it reflects a
distinct physical phenomenon of driver behavior as described above. 

In
Fig. 1 we demonstrate the effect of the slow-to-start rule on traffic flow; 
in particular we
contrast typical `snap-shots' of the steady state with
$P_{slow}=0$ (left panel) and $P_{slow}=0.5$ (right panel). It is clear that
a qualitative change occurs in the distribution as a result of introducing
a non-zero $P_{slow}$;
the queues become less fragmented and the inter-queue regions
widen. In fact, the two rules compete in this respect: the slow-to-start 
rule
causes queues to merge while the disorder nucleates new queues. 
The mean length of queues in the steady state depends critically on the
relative values of $P_{slow}$ and $P_{fault}$. We
shall see later that this interplay can have important consequences for 
highway junctions. 

Figure 2 shows results for the flux $f$ of cars as a function of the
car density $\rho$. The flux is defined as the number of cars moving
in a given step divided by the number of sites, and is therefore a
measure of the highway's efficiency.  
The three flux-density relations 
obtained from the 
CA simulation correspond to
the slow-to-start rule alone (`experiment' -- long-dashed line), 
the disorder rule alone 
(`experiment' -- short-dashed line), and both rules
together (`experiment' -- solid line). Also shown  are the analytical results 
from the theory presented below (`theory' -- solid circles). Note that 
theory and experiment are indistinguishable for the top two curves. 
These plots are for the case $s_{max}=1$, but the theory
developed
below is actually valid for all $s_{max}$.
The CA results are obtained 
from
simulations on a chain of 1500 sites. A periodic boundary condition is
assumed so that both the total number of cars and $\rho$ are conserved.
This is the usual boundary condition for traffic simulation, although
`open' CA models without conservation have also been studied \cite
{Derrida}. For each initial configuration of cars, results are
obtained by averaging over 1000 time steps {\em after} the first 2000
steps, so that the long-time limit is approached. This criterion was found to be
sufficient to guarantee a steady-state being reached. For each car density,
results are averaged over 50 different initial configurations.

For the slow-to-start rule acting alone (long-dashed line in Fig. 2) we have chosen
to set the parameter $P_{slow}=0.5$. The maximum flux 
occurs at $\rho=0.4$.
For $\rho>0.4$, the flux decreases linearly with car density. These
features can be understood by considering the system with a high car
density (e.g. $\rho=0.6$). There will always be substantial queues in
this limit which do not contribute to the flux. We can estimate the
flux if we know the density of cars in the free-flowing regions or,
instead, the average number of sites $n$ associated with each free
moving car. This quantity $n$ is determined by the rate at which cars
leave the head of the queue bounding the free-flowing region on its
immediate left. Consider the basic model ($P_{slow}=0$). A car will
move off from the head of the queue on every time step; there will then
be $(s_{max}+1)$ sites per car in the free flowing region \cite{approx}. 
With $P_{slow}
>0$, some cars wait until the second possible time-step before moving
from the head of the queue. Such a car will effectively occupy 
$(s_{max}+1) + s_{max}= 2s_{max}+1$ sites,
because of the wasted turn when the other free moving cars to its
right all move $s_{max}$ sites further away. The proportion of such slow
cars
is $P_{slow}$, so the average number of sites per free-moving car is
$n = (s_{max}+1)(1-P_{slow}) + (2s_{max}+1)P_{slow}=1+s_{max}(P_{slow}+1)$.
Let $Y$ be the number of
free-moving cars and $X$ be the number of cars involved in queues.
These two quantities are related to the total number of sites $N$ and
number of cars $\rho N$ by 
\begin{eqnarray}
 N = X + [1+s_{max}(1+P_{slow})]Y\nonumber\\ 
\rho N = X + Y. 
\end{eqnarray}
the flux is entirely due to the free moving cars and is given by
\begin{equation}
f\equiv {Y\over N}={1\over(1+P_{slow})}(1-\rho) \ .
\end{equation} 
This
function is valid for sufficiently large $\rho$ so as to produce
queues in the steady state, i.e. all $\rho$ for which $X>0$. The
transition \cite{transition} 
occurs
at a
density $\rho={1\over{1+s_{max}(1+P_{slow})}}$. Below this density, the
absence of
queues means that all cars are free flowing, and $f=s_{max}\rho$. A
complete
description of the flux is  \begin{equation}
f=\cases{s_{max}\rho &for $\rho<{1\over{1+s_{max}(1+P_{slow})}}$\cr 
{1\over(1+P_{slow})}(1-\rho)&for $\rho>{1\over{1+s_{max}(1+P_{slow})}}\
.$\cr}
\end{equation} 
For $P_{slow}=1/2$, the turning point should arise at
$\rho=2/5$, after which the gradient should be $2/3$. This analytic
result matches exactly with the simulation in Fig. 2, where the
peak flux lies at $\rho=0.4$, and the gradient in the region
$\rho>0.4$ is $\approx 0.65$. Such good agreement has also been found
for
other values of $P_{slow}$ and $s_{max}$.
It is interesting to note that strictly in the limit of $P_{fault}=0$ and $s_{max}=1$,
 the
action of the slow-to-start rule becomes identical to the {\em cruise-control} 
rule \cite{cruise}. However, these two rules are {\em not} identical 
for $P_{fault}>0$
or $s_{max}>1$.
We note that the expression in Eq. (3) reduces in the $P_{slow}=0$ limit to the
expression 
\begin{equation} 
f=\cases{s_{max}\rho& for $\rho<(s_{max}+1)^{-1}$\cr
  (1-\rho)& for $\rho>(s_{max}+1)^{-1}\ \ .$\cr} 
\end{equation} 
which has been obtained by Nagel and Herrmann\cite{Nagel2} among others. 

For the disorder rule (short-dashed line in Fig. 2) we have chosen to set
the parameter $P_{fault}=0.1$. 
It turns out that one can
derive an exact  analytic expression for this $s_{max}=1$ system
(see eg. \cite{cluster}). The  form is
\begin{equation}
f={{1-\sqrt{1-4(1-P_{fault})\rho(1-\rho)}}\over2}
\end{equation}
This may be obtained using an `$n$-cluster expansion' in which one
considers
the probabilities $P(c_1,c_2,..c_n)$
of finding a randomly selected string of $n$ consecutive
cells to be in states $c_1,c_2,...c_n$. Whilst the above exact expression
for $s_{max}=1$ can 
be obtained using just an $n=2$ cluster treatment, it was found that 
in order to closely model systems with higher $s_{max}$, correspondingly
larger clusters must be considered. This is understood to result from 
long-range correlations that exist in all systems with $s_{max}>1$. 

The combined action of these two rules produces the `fundamental diagram' shown
as a solid line in Fig. 2,  for a system with $s_{max}=1$ and both $P_{fault}$
and $P_{slow}$ finite. Along this line we display (solid circles) the
 analytic results that we obtained from an $n=2$ cluster treatment (see
the Appendix for an outline derivation).  We can see that the fit, whilst 
good, is no longer exact. This is an indication that introducing 
the slow-to-start rule 
increases the distance over which correlations exist. 
In fact, this correlation can 
be traced to the lengthening of queues beyond the 
statistically expected length (cf. Fig. 1).

We now turn to the highway containing a junction where cars may enter
and leave. Two nearby, but non-adjacent, sites are chosen to be the
`input' and the `output' sites, with the input site to the right of
the output site so that cars entering must essentially traverse the entire
road before exiting. Associated with the input site is an integer $Q$
which is the number of cars queuing in the feeder road (or `ramp')
waiting to enter the highway. Cars are added periodically to the input
queue $(Q \rightarrow Q + 1$); we choose a rate of $1$ car added every
$5$ time steps. Whenever $Q>0$ and the input site is vacant, this site
becomes occupied and $Q\rightarrow Q-1$. We delete one car entering
the output site for every car added to the highway so that the total
number of cars on the road is conserved, apart from small
fluctuations in short time intervals between the addition and removal
of cars. 
The quantity
$Q$ thus measures the flow of cars through the junction. It is
desirable to keep $Q$ low; indeed real junctions may only be able to
support a finite number of waiting cars before becoming
catastrophically locked up. 

Figure 3 shows $\bar{Q}$, the value of $Q$ averaged over the last
$2000$ of $4000$ steps, as a function of the disorder probability
$P_{fault}$ for different values of $s_{max}$ and $P_{slow}$. A car
density of $\rho =0.5$ is chosen for all simulations. The lowest three
lines correspond to $s_{max}=1$ and $P_{slow}=0,0.25,0.5$,
respectively. With $P_{slow}=0$, $\bar{Q}$ is small and increases
slightly with $P_{fault}$. This is expected since in the steady
state of the corresponding junctionless model, every other site
is empty in the $P_{fault}\rightarrow 0$ 
limit. 
The introduction of a
single junction does not significantly alter this distribution, so cars can
easily filter onto the road and $Q(t)$ remains small for all $t$. If we
increase $P_{slow}$ while setting $P_{fault}=0$, $Q(t)$ behaves very
differently, occasionally flaring up to large values. This is shown in
the left inset, in which the number of dots in a column represents the
value of $Q$, and each successive column is advanced by 5 time steps.
The typical value of the maximum does of course depend on the
parameters such as $\rho$, $P_{slow}$, and the rate of adding cars to
the queue in the feeder road; however the appearance of this feature is
quite general.  In order to understand the phenomenon we examine the
spatial distribution of cars in the steady state. We find that the
slow-to-start rule, with or without the junction, gives a steady
state with fewer but longer queues relative to the basic model. When
one of these queues, which move backwards along the road, passes a
junction, it prevents cars from entering the road for a substantial
period of time. It is during this time that the value of $Q$ increases
dramatically. 
It is worth spending a moment looking into why such large queues form
on roads (with or without junctions)
employing the slow-to-start rule alone (no disorder, i.e. $P_{fault}=0$). 
The explanation lies not with the
fact that the cars are slow to move off, but rather with the
uncertainty in this delay, which allows the queue lengths to vary. It
is possible for a queue to shrink to zero, i.e. evaporate completely,
but there is no corresponding mechanism allowing for creation of
queues. Since the total number of queued cars remains approximately
constant,  it is clear that the average length will
increase. The ultimate limit for a closed system is that all queues
merge into one; however it would take an astronomical time to reach
this state on a long road.

If we now allow both $P_{fault}$ and $P_{slow}$ to be non-zero, we
make the interesting observation that {\em disorder} can {\em improve}
the junction's performance. The third curve in Fig. 3 drops dramatically
as $P_{fault}$ is increased from zero. The value of $\bar{Q}$ is
reduced from $1.9$ for $P_{fault} = 0$ to $0.5$ for $P_{fault} =
0.025$. In the corresponding inset, we see that the value of $Q$ no
longer flares up. The disorder rule breaks up the long queues
resulting from the slow-to-start rule alone by increasing the {\em
number} of queues, without significantly altering the car density in
the inter-queue regions. Each momentary driving defect has a chance of
nucleating a queue. As the average length of the queues on the road
decreases, the maximum number of cars waiting to enter also decreases. 

The surprising conclusion from the model is that it is `beneficial' to create
queues on the highway. The queues effectively compete with one another
for the static cars, the total number of which remains practically constant
over time. When a queue becomes deprived of static
cars it is destroyed, so without the introduction of new queues the
road becomes highly inhomogeneous: only a small number of large queues
survive. This inhomogeneity has a markedly detrimental effect on road
junctions, which seize up when one of the large queues moves past.

Finally we turn to the upper three lines in Fig. 3, which between them
display the effect of increasing $s_{max}$ at a fixed value of
$P_{slow}=0.5$. We observe that with increasing $s_{max}$, (a) $\bar{Q}$
increases, and (b) the beneficial effect of increasing $P_{fault}$
diminishes. To understand observation (a), we note that the queue
lengths 
along the road (which directly affect $\bar{Q}$) increase with the 
choice of $s_{max}$; in fact this occurs
for any given values of $P_{fault}$ and $P_{slow}$, so we may take the
limit 
$P_{fault}=P_{slow}=0$ to understand the effect.
Since 
for 
$\rho\ge0.5$ the flux is independent of $s_{max}$ (see
Eq. (4))
 it is clear 
that as $s_{max}$ increases, the cars traveling at high speed must be 
counter-balanced 
by a greater number stuck in queues. 
Observation (b) is due to the inability of the disorder rule to
nucleate queues when it acts on fast-moving cars. For a queue to form
spontaneously, a car must be made stationary. However the disorder
rule only reduces the velocity of cars by one unit, so that the
probability of actually halting a car which initially moves with speed
$s_{max}$ (by the disorder rule, this would imply decelerating it on
$s_{max}$ consecutive  time-steps) falls approximately as
$P_{fault}$ to the power of $s_{max}$. In real traffic situations, it
is indeed the case that jams do not tend to form spontaneously in
regions of a road where cars are moving very quickly.

In conclusion, we have studied the performance of a junction under the
action of three different rules. The acceleration and disorder rules
due to Nagel and Schrekenberg were employed, and we introduced a third {\em
slow-to-start} rule which reflects a feature of real driving that is
distinct from general disorder. Having quoted and obtained analytic
forms describing the effects of these rules on a bare road, we
then applied them to the junction. We measured the junction's
performance by the variable $Q$, which is the length of the queue of
cars forced to wait  in the `feeder-road' or `ramp'. We studied in detail the
effects when the road is half-filled with cars, i.e. $\rho = 0.5$. Our
main findings, for which we have provided qualitative explanations, are
as follows: (i) when the cars on a highway are constrained to move
slowly (i.e. $s_{max}=1$) the junction's performance is maximized by
introducing a finite disordering along the road. (ii) As the speed limit
is relaxed, i.e. for larger $s_{max}$, we note that (a) the
performance of the junction is reduced, and  (b) the beneficial effect
of disorder diminishes. Noting that for $\rho\geq0.5$ the flux along the
 road
is not significantly altered by the choice of $s_{max}$, we may 
conclude that
{\em it is desirable to set a speed limit near junctions on busy
single-lane roads}. 

The systems we have studied were designed
to be both plausible and intuitive and yet still permit a certain degree of
analytical analysis. We believe that the general characteristics
of our model are indeed consistent with  personal experience. In
order to establish the extent to which this (or any) CA traffic model
makes accurate {\em quantitative} predictions about real traffic flow, one
should clearly make a thorough comparison with empirical traffic data for the
same highway/junction system, if available. Such a
comparison lies beyond the intended scope of this paper.

This study was supported by an EPSRC Studentship (S.C.B.) and the
Nuffield Foundation (N.F.J.). Work at the Chinese University of Hong
Kong was supported in part by a Direct Grant for Research 1994-95. One of
us 
(P.M.H.) is a member of a research team on traffic problems in modern
cities 
supported by the Shanghai Natural Science Foundation, Shanghai, China.

\newpage

{\bf APPENDIX}

Here we outline the method for applying the cluster
expansion
\cite{cluster} to our model with $s_{max}=1$ and finite $P_{slow}$. When we
consider the state of the array just after movement, we see that there are
four states  that a cell may be in: car
moving with velocity 1 (denoted `1'); car static due to the 
action of the slow-to-start or disorder rules (denoted `s'); car
static due to blockage ahead 
(denoted `b'); and an empty cell (denoted `e'). Following \cite{cluster}, we 
can reduce the problem to three possible states by considering the road at an 
intermediate stage between one movement and the next. 
Consider a single time-step
as consisting of the following stages: (i) acceleration --
all cars are assigned a velocity of `1'; (ii) slow-to-start -- all cars that
are 
legitimate
candidates may be decelerated to `s' with probability $P_{slow}$; (iii)
blockage -- all
cars which are blocked from moving have their state changed from `1' to `b';
(iv) 
disorder -- all cars still in state `1' may be decelerated to `s' with
probability 
$P_{fault}$; (v) movement -- all cars in state `1' are moved one cell to
the right. Now 
consider the state of the road
after the action of the acceleration and slow-to-start stages, but
before 
the action of the blockage stage. In this way we
avoid having to consider cells in state         `b'. 
When we come to the expression
for the flux we must apply the remaining rules, i.e. consider only  $P(10)$ and
apply the factor $(1-P_{fault})$. 
Recall that $P(c_1,c_2)$
is the probability of finding a randomly selected string of 2 consecutive cells
with states $c_1$ and $c_2$ respectively. In the present work we consider only a
2-cluster expansion; this was found to
be sufficient to model the corresponding fundamental diagram to within 2\%
accuracy (see  Fig. 2). The
2-cluster probabilities obey the following identities at all times:
\begin{eqnarray}
P(S1)=P(SS)=0\ \ \ \ \ P(S0)&=&P(1S)+P(0S) \nonumber \\
P(01)&=&P(10)+P(1S) \nonumber \\
\rho&=&P(01)+P(11)+P(S0) \\
1-\rho&=&P(00)+P(10)+P(S0) \nonumber 
\end{eqnarray}
With the application of these identities we are left requiring three
equations. We employ the following, which are approximately true in the
equilibrium limit:
\begin{eqnarray}
P(S0)&=&q(1-s)(P(1110)+P(0110)) \nonumber \\
P(11)&=&\epsilon(qP(1011)+ P(1111)+  P(0111))+ P(11S0)+ P(111S)+ qP(10S0)+
\nonumber \\ 
     & &P(01S0)+ P(011S)+ pP(1110)+   p(P0110)+ q(P101S)+ qpP(1010)
\nonumber \\ 
P(10)&=&pP(0100)+ qsP(0110)+  qP(1000)+ q^2P(1010)+ 
P(1S00)+ P(0S00)+ \nonumber \\
     & &qP(1001)+ qsP(1110)+  pP(0101)+ pP(1100)+ pP(1101)+ P(1S01)+
 \nonumber \\
     & &P(0S01)+ qP(100S)+ pP(010S)+ pP(110S)+ P(0S0S)+ P(1S0S)
\end{eqnarray}
For compactness we have used
\begin{equation}
p=P_{fault}\ \ \ \ q=(1-P_{fault})\ \ \ \ s=1-P_{slow}
\end{equation}
The quantity $P(WXYZ)$ is of course expanded in the 2-cluster expansion as
\begin{equation}
P(WXYZ)=P(W|\bar X)P(XY)P(\bar Y|Z)
\end{equation}
where the conditional probabilities
\begin{equation}
P(W|\bar X)={{P(WX)}\over{\sum_i P(Wi)}}\ \ \ \ 
P(\bar Y|Z)={{P(YZ)}\over{\sum_i P(iZ)}}
\end{equation}
Note the important correction $\epsilon$ in the expression for $P(11)$:
\begin{equation}
\epsilon=(1-qpP(\bar 1|1))
\end{equation}
The seven equations contained in expressions (7) and (8) are solved
simultaneously to find the quantity
$P(10)$  in terms of the constants $\rho,P_{fault},P_{slow}$. The flux then
follows
by  $f=qP(10)$.

\newpage \centerline{\bf Figure Captions}

\bigskip

\noindent Figure 1: Each panel displays the distribution of cars along a
highway over 500 consecutive time steps; a black pixel represents a car whilst a
white pixel corresponds to an empty cell. The section of road is 400 cells wide.
Left panel: $P_{fault}=0.25$, $P_{slow}=0$; right panel: $P_{fault}=0.25$,
$P_{slow}=0.5$. 
Typical parameter values for a realistic highway are expected to 
lie between these two cases. For both panels $s_{max}=3$.

\bigskip

\noindent Figure 2: Flux-density relations (i.e. `fundamental diagram') for
various values of $P_{fault}$ and $P_{slow}$. In all cases $s_{max}=1$. The
CA simulation results are shown for the slow-to-start rule (long-dashed line),
the disorder rule (short-dashed line) and the combination of the two rules
(solid line). Also shown are the analytic results (solid circles). Note that for
the upper two lines, the analytic results (`theory') and the CA
simulation results (`experiment') are indistinguishable.

\bigskip

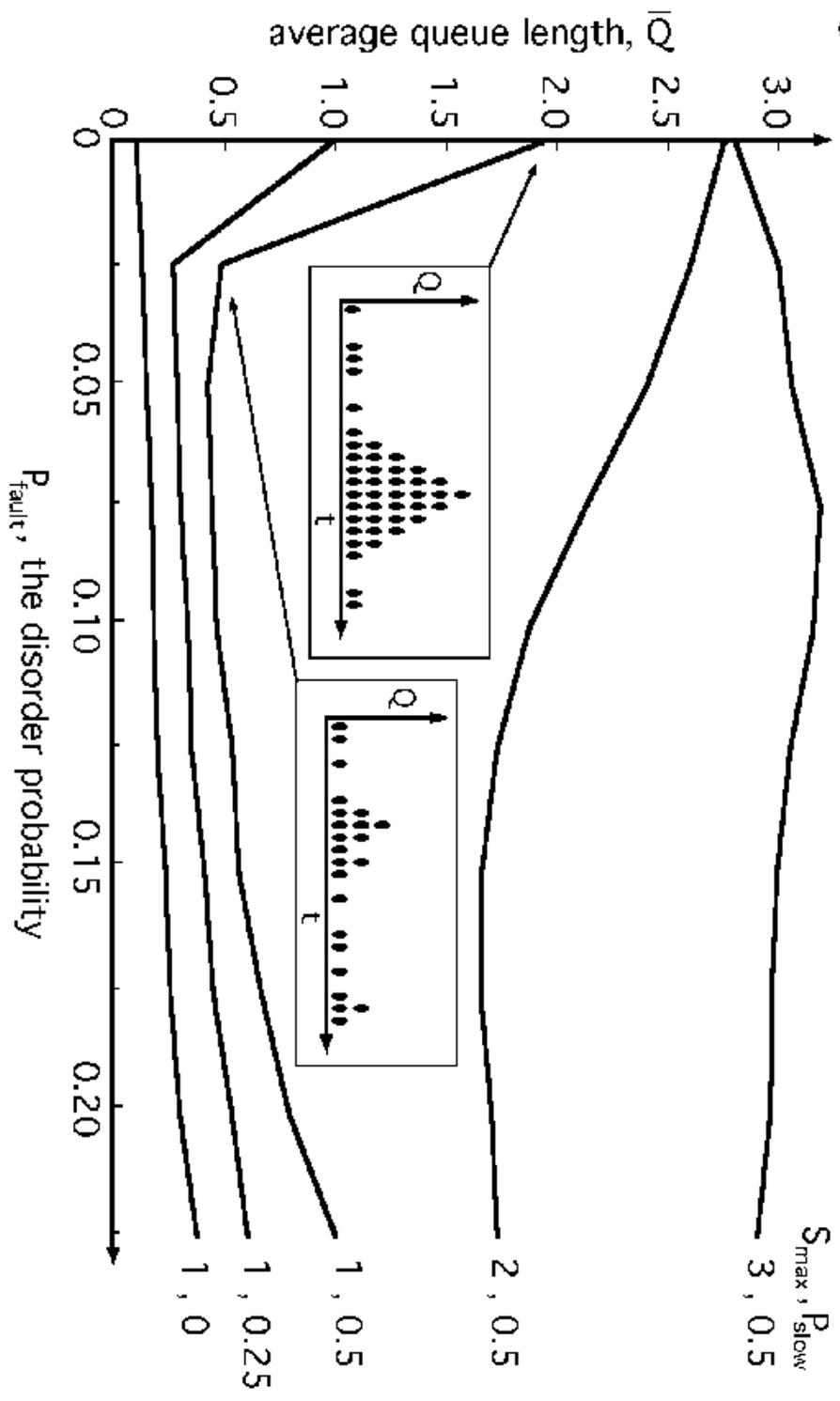
\noindent Figure 3: Average length of the feeder-road queue $\bar{Q}$ as
a function of $P_{fault}$ (disorder rule) for various values of
$s_{max}$ (acceleration rule) and $P_{slow}$ (slow-to-start rule).
Insets: the number of dots in each column represents $Q$ at a given time-step
 and each successive column represents an advance of 5 time-steps.


\begin{thebibliography}{99}
\bibitem{transition} If $\rho$ lies in the range
${1\over{1+s_{max}(1+P_{slow})}}$ to ${1\over{s_{max}+1}}$, then a certain
small
sub-set of initial configurations will result in a steady-state with no
traffic
queues.  The system therefore has two possible steady state solutions for
any
$\rho$ in this region; however the queue-less  solution will tend to be
`washed out'
when we average over many  initial configurations; moreover, it will
vanish
entirely when we  move to the  more realistic $P_{fault}>0$ model.
\end{thebibliography}
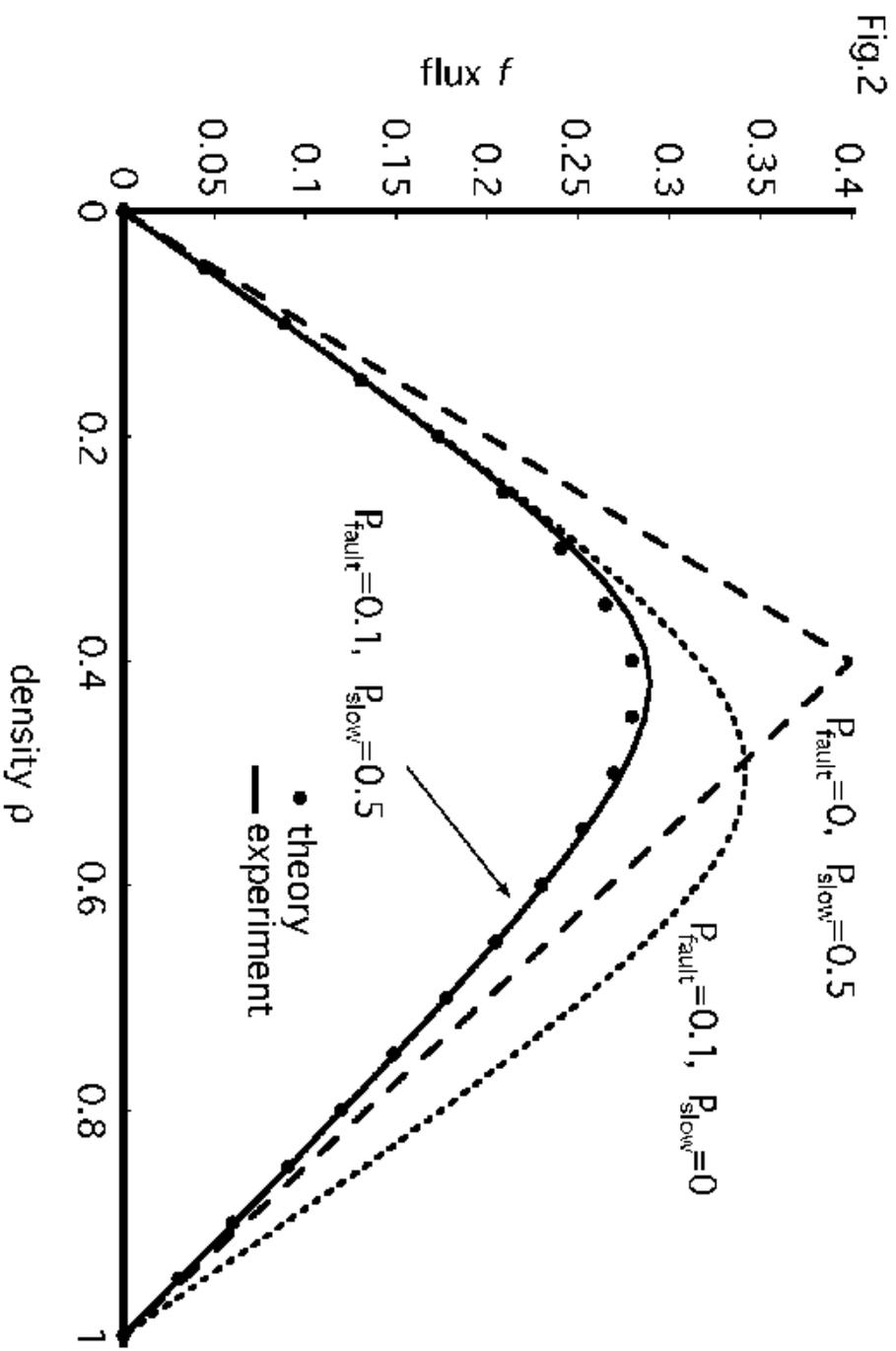
\end{document}